\documentclass[aps,pra,twocolumn,superscriptaddress,nofootinbib]{revtex4-1}
\usepackage{amsmath}
\usepackage{amssymb}
\usepackage{graphicx}
\usepackage{mathrsfs}
\usepackage{ntheorem}
\usepackage{mathtools}
\usepackage{enumerate}
\usepackage{enumitem}
\usepackage{times,txfonts}
\usepackage{subfigure}
\usepackage{upgreek}
\usepackage{tikz}
\newcommand{\ket}[1]{|#1\rangle}
\newcommand{\bra}[1]{\langle #1|}
\newcommand{\Tr}{\mathrm{Tr}}

\def\CC{{\rm\kern.24em \vrule width.04em height1.46ex depth-.07ex \kern-.30em C}}
\def\RR{{\rm\kern.24em \vrule width.04em height1.46ex depth-.07ex
\kern-.30em R}}
\def\P{{\rm I\kern-.25em P}}

\begin{document}

\title{Examining the validity of Schatten-$p$-norm-based functionals as coherence measures}

\author{Xiao-Dan Cui}
\address{Department of Physics, Shandong University, Jinan 250100, China}

\author{C. L. Liu}
\email{clliusdu@foxmail.com}
\affiliation{Department of Physics, Shandong University, Jinan 250100, China}
\affiliation{Institute of Physics, Beijing National Laboratory for
  Condensed Matter Physics, Chinese Academy of Sciences, Beijing
  100190, China}

\author{D. M. Tong}
\email{tdm@sdu.edu.cn}
\affiliation{Department of Physics, Shandong University, Jinan 250100, China}
\date{\today}

\begin{abstract}
It has been asked by different authors whether the two classes of Schatten-$p$-norm-based functionals $C_p(\rho)=\min_{\sigma\in\mathcal{I}}||\rho-\sigma||_p$ and $ \tilde{C}_p(\rho)= \|\rho-\Delta\rho\|_{p}$  with $p\geq 1$ are valid coherence measures under incoherent operations, strictly incoherent operations, and genuinely incoherent operations, respectively, where $\mathcal{I}$ is the set of incoherent states and $\Delta\rho$ is the diagonal part of density operator $\rho$. Of these questions, all we know is that $C_p(\rho)$ is not a valid coherence measure under incoherent operations and strictly incoherent operations, but all other aspects remain open. In this paper, we prove that (1) $\tilde{C}_1(\rho)$ is a valid coherence measure under both strictly incoherent operations and genuinely incoherent operations  but not a valid coherence measure under incoherent operations,  (2) $C_1(\rho)$ is not a valid coherence measure even under genuinely incoherent operations, and (3) neither ${C}_{p>1}(\rho)$ nor $\tilde{C}_{p>1}(\rho)$ is a valid coherence measure under any of the three sets of operations. This paper not only  provides a thorough examination on the validity of taking $C_p(\rho)$ and $\tilde{C}_p(\rho)$ as coherence measures, but also finds an example that fulfills the monotonicity under strictly incoherent operations but violates it under incoherent operations.
\end{abstract}
\maketitle

\section{Introduction}
Quantum coherence is a fundamental property of quantum mechanics and describes the capability of a quantum state to exhibit quantum interference phenomena. It provides an important resource for various quantum information processing tasks, such as quantum algorithms,  quantum cryptography \cite{Nielsen}, nanoscale thermodynamics \cite{Lostaglio1}, quantum metrology \cite{Giovannetti}, and quantum biology \cite{Lloyd,Huelga}. The resource theory of coherence has attracted a growing interest due to the rapid development of quantum information science \cite{Aberg,Baumgratz,Chitambar,Chitambar1,Yadin,Vicente,Fan,Streltsov,Levi}.

All quantum resource theories have two fundamental ingredients: free states and free operations \cite{Liuz,Coecke,Horodecki,Brandao,Chitambar2,Liu2}. For the resource theory of coherence, the free states are the quantum states that are diagonal in a prefixed reference basis. However, there is no general consensus on the definition of free operations. With different physical and mathematical considerations, researchers have proposed a number of free operations, such as maximally incoherent operations \cite{Aberg}, incoherent operations \cite{Baumgratz,Levi}, strictly incoherent operations \cite{Yadin}, genuinely incoherent operations \cite{Vicente}, and others \cite{Vicente,Chitambar,Chitambar1}. Each definition of free operations has its own characteristics. For instance, incoherent operations can never generate, not even probabilistically, coherence from an incoherent state \cite{Baumgratz}; strictly incoherent operations can neither create nor use coherence and have a physical interpretation in terms of interferometry  \cite{Yadin}; while the genuinely incoherent operations can capture coherence under additional constraints such as energy preservation \cite{Vicente}.

Based on the notions of incoherent states and free operations, frameworks for quantifying coherence can be established. A framework for the quantification of coherence usually consists of four conditions: (C1) the coherence being zero (positive) for incoherent states (all other states), (C2) the monotonicity of coherence under free operations of coherence, (C3) the monotonicity of coherence under selective measurements on average, and (C4) the nonincreasing of coherence under mixing of quantum states. A functional of density operators can be taken as a valid coherence measure if and only if it satisfies the four conditions.

By following the four conditions of quantifying coherence, a number of coherence measures, such as the $l_1$ norm of coherence \cite{Baumgratz}, the relative entropy of coherence \cite{Baumgratz}, the robustness of coherence \cite{Napoli,Piani}, and the coherence of formation \cite{Aberg,Yuan,Winter}, have been proposed, and these measures have been widely used to address various topics on quantum coherence \cite{Bromley,Yu1,Bu,Liu,Regula1,Xi,Streltsov1,Yao,Radhakrishnan,Guo,Tan,Mat,Liu3,Singh,Ma,Chitambar4,Chitambar3,Streltsov3,Regula,Zhangc,Yu,Zhang,Liu1,Liu5,Winter,Yuan}, such as the dynamics of quantum coherence \cite{Bromley,Yu1,Bu}, the distillation of quantum coherence \cite{Winter,Liu,Regula1,Yuan}, and the relations between quantum coherence and other quantum resources \cite{Streltsov1,Radhakrishnan,Yao,Xi,Ma,Tan,Chitambar4,Guo,Singh,Chitambar3,Radhakrishnan,Yao,Streltsov3,Regula,Liu3,Mat,Zhangc}.

Clearly, whether a functional of density operators can be taken as a valid coherence measure is also dependent on the free operations taken, besides the property of the functional itself. Since there are different definitions of free operations, a functional being a valid coherence measure under one set of free operations may not be valid under another set unless the latter belongs to the former  as a subset. For example, all the functionals mentioned above are valid coherence measures under incoherent operations, strictly incoherent operations, and genuinely incoherent operations, but the $l_1$ norm of coherence and the coherence of formation are not valid coherence measures under maximally incoherent operations \cite{Hu,Bu3}. However, so far, we have not been able to find any examples that are measures under strictly incoherent operations but not measures under incoherent operations. This issue was first raised in Ref. \cite{Yadin}.

In this paper, we examine two classes of Schatten-$p$-norm-based functionals
\begin{eqnarray}
C_p(\rho)=\min_{\sigma\in\mathcal{I}}||\rho-\sigma||_p \label{measure'}
\end{eqnarray}
and
\begin{eqnarray}
 \tilde{C}_p(\rho)= \|\rho-\Delta\rho\|_{p} \label{measure}
\end{eqnarray}
with $p\geq 1$, which were proposed in Refs. \cite{Baumgratz} and \cite{Vicente}, respectively.
Here, the Schatten-$p$ norm $\|\cdot\|_{p}$ is defined as
 \begin{eqnarray}
\|M\|_{p}=\left(\mathrm{Tr}[(M^\dag M)^{\frac{p}{2}}]\right)^{\frac{1}{p}},
  \end{eqnarray}
$\mathcal{I}$ represents the set of incoherent states, and  $\Delta\rho=\sum_i\ket{i}\bra{i}\rho\ket{i}\bra{i}$ is the diagonal part of density operator $\rho$.

It has been asked by different authors whether the Schatten-$p$-norm-based functionals are valid coherence measures under incoherent operations, strictly incoherent operations, and genuinely incoherent operations, respectively. Several previous papers have addressed these questions \cite{Baumgratz,Yu,Rana,Vicente,Shao}. However, it is a nontrivial work to answer them. For example, $C_1(\rho)$, i.e., $C_p(\rho)$ for $p=1$, had been expected as a valid coherence measure under incoherent operations,  but it finally proved invalid by the authors in Ref. \cite{Yu} after several efforts made by others \cite{Shao,Rana}. Also, $C_{p>1}(\rho)$ has been proved not to be a valid coherence measure under both incoherent operations and strictly incoherent operations in Ref. \cite{Rana} but it is unclear whether $C_{p>1}(\rho)$ is a valid coherence measure or not under genuinely incoherent operations. $\tilde{C}_p(\rho)$ has been proved to satisfy conditions (C1), (C2), and (C4) under genuinely incoherent operations \cite{Vicente} but has not been proved to satisfy condition (C3) or not, and therefore it is not clear whether $\tilde{C}_p(\rho)$ is a valid coherence measure under genuinely incoherent operations, not to mention it under incoherent operations or strictly incoherent operations. So far, about whether the two classes of Schatten-$p$-norm-based functionals can be taken as valid coherence measures under incoherent operations, strictly incoherent operations, and genuinely incoherent operations, all we know is that $C_p(\rho)$ is not a valid coherence measure under incoherent operations and strictly incoherent operations, but all other aspects remain open, as shown in Table I. In this paper, we will resolve these open questions, filling up the gaps in the table.

 \begin{table*}\label{table1}
\caption{Of whether $C_p(\rho)$ and $\tilde{C}_p(\rho)$  can be taken as coherence measures  under incoherent operations, strictly incoherent operations, and genuinely incoherent operations, all we know is that $C_p(\rho)$ is not a valid coherence measure under incoherent operations and strictly incoherent operations, but all other aspects remain unknown.}
\begin{tabular}{cccc}
    \hline
    \hline
             &  Incoherent operations    &  Strictly incoherent operations  & Genuinely incoherent operations  \\
    \hline
    $C_1(\rho)$ & Not a coherence measure& Not a coherence measure & Unknown \\
    $\tilde{C}_1(\rho)$ & Unknown & Unknown& Unknown\\
    $C_{p>1}(\rho)$ &  Not a coherence measure& Not a coherence measure   & Unknown \\
    $\tilde{C}_{p>1}(\rho)$ & Unknown & Unknown & Unknown  \\
    \hline
    \hline
    \end{tabular}
\end{table*}

The paper is organized as follows. In Sec. II, we review some notions related to coherence measures. In Sec. III, we present one by one our main results on whether $C_p(\rho)$ and $ \tilde{C}_p(\rho)$ can be taken as coherence measures.  Section IV presents our conclusions.

\section{Preliminaries}
We recapitulate some notions related to coherence measures, such as incoherent states, incoherent operations, strictly incoherent operations, genuinely incoherent operations, and frameworks of quantifying coherence.

Let $\mathcal{H}$ represent the Hilbert space of a $d$-dimensional quantum system. A particular basis of $\mathcal{H}$ is denoted as $\{\ket{i},~i=0,1,...,d-1\}$, known as the incoherent basis, which is chosen according to the physical problem under consideration. The coherence of a state is then measured based on the basis chosen. We use $\rho=\sum_{ij}\rho_{ij}\ket{i}\bra{j}$  to denote a general density operator in the basis, where  $\rho_{ij}$ are the elements of the density matrix.  A state is called an incoherent state, specially denoted as $\sigma$,  if its density operator is diagonal in the basis, and the set of all incoherent states is denoted by $\mathcal {I}$. It follows that a density operator $\rho$ belonging to $\mathcal {I}$ is of the form
\begin{eqnarray}
\sigma=\sum^{d-1}_{i=0}\sigma_{ii}\ket{i}\bra{i}.
\end{eqnarray}
All other states which cannot be written as diagonal matrices in the basis are called coherent states.

An incoherent operation \cite{Baumgratz} is defined by a completely positive and trace preserving  map,
\begin{eqnarray}\label{Tong1}
\Lambda(\rho)=\sum_n K_n\rho K_n^\dagger
\end{eqnarray}
 with the Kraus operators fulfilling not only
$\sum_n K_n^\dagger K_n= I$ but also
\begin{eqnarray}\label{Tong2}
K_n\mathcal{I}K_n^\dagger\subset \mathcal{I},
\end{eqnarray}
i.e., each $K_n$ maps an incoherent state to an incoherent state.

A strictly incoherent operation \cite{Yadin,Winter} is defined by a completely positive and trace preserving  map, $\Lambda(\rho)=\sum_n K_n\rho K_n^\dagger$ with the Kraus operators fulfilling not only $\sum_n K_n^\dagger K_n= I$  and $K_n\mathcal{I}K_n^\dagger\subset \mathcal{I}$ but also
\begin{eqnarray}\label{Tong3}
K_n^\dagger\mathcal{I}K_n\subset\mathcal{I},
\end{eqnarray}
i.e., each $K_n$ as well $K_n^\dagger$ maps an incoherent state to an incoherent state.

A genuinely incoherent operation \cite{Vicente} is defined by a completely positive and trace preserving map, $\Lambda(\rho)=\sum_n K_n\rho K_n^\dagger$ with the Kraus operators fulfilling not only $\sum_n K_n^\dagger K_n= I$ but also
\begin{eqnarray}
 \Lambda_G(\sigma)=\sum_n K_n \sigma K_n^\dagger=\sigma,
\end{eqnarray}
i.e., all incoherent states are fixed points and therefore each $K_n$ must be diagonal. Obviously, the Kraus operators of a genuinely incoherent operation naturally fulfill Eqs. (\ref{Tong2}) and (\ref{Tong3}) too.

If we use $\mathcal{S}_{IO}$, $\mathcal{S}_{SIO}$, and $\mathcal{S}_{GIO}$ to represent the sets of incoherent operations, strictly incoherent operations, and genuinely incoherent operations, respectively, they have the inclusion relationships
\begin{eqnarray}
\mathcal{S}_{GIO}\subset \mathcal{S}_{SIO} \subset \mathcal{S}_{IO}.
\end{eqnarray}

A framework for quantifying coherence usually consists of four conditions \cite{Baumgratz}.  A functional $C$ can be taken as a valid coherence measure, if it satisfies the following four conditions:
\begin{itemize}
\item[(C1)] $C(\rho)\ge 0$, and $C(\rho)=0$ if and only if $\rho\in\mathcal{I}$;
\item[(C2)] It has monotonicity under free operations: $C(\rho)\ge C(\Lambda(\rho))$ if $\Lambda$ is a free operation;
\item[(C3)] It has monotonicity under selective measurement on average:
\begin{eqnarray}
C(\rho)\ge \sum_np_nC(\rho_n), \label{C3}
\end{eqnarray}
 where $p_n=\Tr(K_n\rho K_n^\dagger)$,
$\rho_n=K_n\rho K_n^\dagger/p_n$, and $\Lambda$ is a free operation;
\item[(C4)] It is nonincreasing under mixing of quantum states, i.e., convexity: $\sum_nq_nC(\rho_n)\ge C(\sum_nq_n\rho_n)$ for any set of states $\{\rho_n\}$ and any probability distribution $\{q_n\}$.
\end{itemize}

An alternative statement of the framework for quantifying coherence consists of three conditions \cite{Yu}, instead of the above four conditions. A functional $C$ can be taken as a valid coherence measure, if it satisfies the following three conditions:
\begin{itemize}
\item[(A1)] $C(\rho)\geq 0$ for all states, and $C(\rho)=0$ if and only if $\rho$ are incoherent states;
\item[(A2)] $C(\rho)\geq C(\Lambda(\rho))$ if $\Lambda$ is a free operation;
\item[(A3)] For all block-diagonal states $\rho$ in the incoherent basis, there is
$C(p_1\rho_1\oplus p_2\rho_2)=p_1C(\rho_1)+p_2C(\rho_2)$.
\end{itemize}

The above two statements are exactly equivalent under incoherent operations and strictly incoherent operations, and hence one can use any of them to examine whether a functional $C(\rho)$ can be taken as a coherence measure.

\section{Main Results}

With these preliminaries, we may now present our results.  We will first show that $\tilde{C}_1(\rho)$ is a valid coherence measure under strictly incoherent operations and genuinely incoherent operations but is not a valid coherence measure under  incoherent operations, and we then demonstrate $C_1(\rho)$, $C_{p>1}(\rho)$, and $ \tilde{C}_{p>1}(\rho)$ can not be taken as coherence measures even under genuinely incoherent operations.

\subsection{$\tilde{C}_1(\rho)$ is a coherence measure under strictly incoherent operations and genuinely incoherent operations}

We first show that $ \tilde{C}_1(\rho)= \|\rho-\Delta\rho\|_{1}$ is a valid coherence measure under strictly incoherent operations. For this, we only need to prove that it fulfills the three conditions $(A1)$, (A2), and (A3).

First, it is straightforward to see  $\tilde{C}_1(\rho)=0$ for $\rho$ being an incoherent state, since $\rho=\Delta\rho$. For $\rho$ being a coherent state, $\sqrt{(\rho-\Delta\rho)^2}$ is a positive semidefinite matrix and there is   $\tilde{C}_1(\rho)=\|\rho-\Delta\rho\|_{1}=\mathrm{Tr}\sqrt{(\rho-\Delta\rho)^2}=\sum_{i}\uplambda_{i}> 0$, where $\uplambda_{i}$ are eigenvalues of $\sqrt{(\rho-\Delta\rho)^2}$ . That is, condition $(A1)$ is valid for  $\tilde{C}_1(\rho)$.

Second, since $\Lambda(\Delta\rho)=\Delta(\Lambda(\rho))$ for any strictly incoherent operation, there is
\begin{align}
\tilde{C}_1(\Lambda(\rho))=\|\Lambda(\rho)-\Delta\left(\Lambda(\rho)\right)\|_{1}
=\|\Lambda(\rho)-\Lambda(\Delta\rho)\|_{1}.
\end{align}
Noting that $\|\rho-\sigma\|_{1}$, as a distance functional, is contractive under trace preserving quantum operations \cite{Nielsen}, satisfying $\|\Lambda(\rho)-\Lambda(\sigma)\|_{1}\leq \|\rho-\sigma\|_1$, we then have
\begin{align}
\tilde{C}_1(\Lambda(\rho))\leq \|\rho-\Delta\rho\|_{1}=\tilde{C}_1(\rho),
\end{align}
i.e., condition $(A2)$ is valid for $\tilde{C}_1(\rho)$.

Third, we show that $\tilde{C}_1(\rho)$ satisfies condition $(A3)$ too.
By the definitions, $\tilde{C}_1(\rho)=\|\rho-\Delta\rho\|_{1}$ and   $\|M\|_1=\mathrm{Tr}\sqrt{M^\dag M}$, we have
\begin{align}\label{Tong4}
\tilde{C}_1(p_{1}\rho_{1}\oplus p_{2}\rho_{2})
&=\mathrm{Tr}\sqrt{\left(p_{1}(\rho_{1}-\Delta\rho_{1})\oplus p_{2}(\rho_{2}-\Delta\rho_{2})\right)^{2}}
\nonumber\\
&=\mathrm{Tr}\sqrt{p_{1}^{2}(\rho_{1}-\Delta\rho_{1})^{2}\oplus p_{2}^{2}(\rho_{2}-\Delta\rho_{2})^{2}}.
\end{align}

To make further calculations, we need to rewrite $\sqrt{p_{1}^{2}(\rho_{1}-\Delta\rho_{1})^{2}\oplus p_{2}^{2}(\rho_{2}-\Delta\rho_{2})^{2}}$ as $\sqrt{p_{1}^{2}(\rho_{1}-\Delta\rho_{1})^{2}}\oplus
\sqrt{p_{2}^{2}(\rho_{2}-\Delta\rho_{2})^{2}}$. The validity of this rewriting can be proved by using the decomposition expression of a Hermitian operator with its eigenvalues and eigenvectors. Indeed, $p_{1}^{2}(\rho_{1}-\Delta\rho_{1})^{2}$ and  $p_{2}^{2}(\rho_{2}-\Delta\rho_{2})^{2}$ are two Hermitian operators, and they can be expressed as $\sum_i\uplambda_{i}|\psi_{i}\rangle\langle\psi_{i}|$ and $\sum_j\uplambda^\prime_j|\psi^\prime_{j}\rangle\langle\psi^\prime_{j}|$, respectively, where $\uplambda_{i}$ and $|\psi_{i}\rangle$  are the eigenvalues and eigenvectors of the first operator, and $\uplambda^\prime_{j}$ and $|\psi^\prime_{j}\rangle$ are the eigenvalues and eigenvectors of the second operator. Then, $\uplambda_{i}$ and $\uplambda^\prime_{j}$ will be the eigenvalues of $p_{1}^{2}(\rho_{1}-\Delta\rho_{1})^{2}\oplus p_{2}^{2}(\rho_{2}-\Delta\rho_{2})^{2}$, and $|\Psi_{i}\rangle=\ket{\psi_i}+|\mathbf{0}\rangle$ and $|\Psi^\prime_{j}\rangle=|\mathbf{0^\prime}\rangle+|\psi^\prime_{j}\rangle$ will be the corresponding eigenvectors of $p_{1}^{2}(\rho_{1}-\Delta\rho_{1})^{2}\oplus p_{2}^{2}(\rho_{2}-\Delta\rho_{2})^{2}$, where  $|\mathbf{0}\rangle$ and  $|\mathbf{0^\prime}\rangle$  are null vectors. Thus, there is
\begin{align}\label{Tong5}
&\sqrt{p_{1}^{2}(\rho_{1}-\Delta\rho_{1})^{2}\oplus p_{2}^{2}(\rho_{2}-\Delta\rho_{2})^{2}}\nonumber\\
&=\sqrt{\sum_{i}\uplambda_{i}|\Psi_{i}\rangle\langle\Psi_{i}|
+\sum_{j}\uplambda^\prime_j|\Psi^\prime_{j}\rangle\langle\Psi^\prime_{j}|}
\nonumber\\
&=\sum_{i}\sqrt{\uplambda_{i}}|\Psi_{i}\rangle\langle\Psi_{i}|
+\sum_{j}\sqrt{\uplambda^\prime_j}|\Psi^\prime_{j}\rangle\langle\Psi^\prime_{j}|
\nonumber\\
&=\sum_{i}\sqrt{\uplambda_{i}}|\psi_{i}\rangle\langle\psi_{i}|
\oplus\sum_{j}\sqrt{\uplambda^\prime_j}|\psi^\prime_{j}\rangle\langle\psi^\prime_{j}|
\nonumber\\
&=\sqrt{p_{1}^{2}(\rho_{1}-\Delta\rho_{1})^{2}}\oplus
\sqrt{p_{2}^{2}(\rho_{2}-\Delta\rho_{2})^{2}}.
\end{align}

Substituting  Eq. (\ref{Tong5}) into Eq. (\ref{Tong4}), we finally obtain
\begin{align}
\tilde{C}_1(p_{1}\rho_{1}\oplus p_{2}\rho_{2})
&=\mathrm{Tr}\sqrt{p_{1}^{2}(\rho_{1}-\Delta\rho_{1})^{2}\oplus p_{2}^{2}(\rho_{2}-\Delta\rho_{2})^{2}}
\nonumber\\
&=\mathrm{Tr}\left(p_{1}\sqrt{\left(\rho_{1}-\Delta\rho_{1}\right)^{2}}\oplus p_{2}\sqrt{\left(\rho_{2}-\Delta\rho_{2}\right)^{2}}\right)
\nonumber\\
&=p_{1}\tilde{C}_1(\rho_{1})+p_{2}\tilde{C}_1(\rho_{2}), \label{direct_sum}
\end{align}
i.e., $\tilde{C}_1(\rho)$ satisfies condition $(A3)$.

The above discussion shows that $\tilde{C}_1(\rho)$ fulfills all the three conditions $(A1)$, $(A2)$, and $(A3)$, and hence it is a valid coherence measure under strictly incoherent operations. Certainly, it is also a valid coherence measure under genuinely incoherent operations, since the set of genuinely incoherent operations is a subset of strictly incoherent operations.

\subsection{$\tilde{C}_1(\rho)$ is not a coherence measure under incoherent operations}

Although $\tilde{C}_1(\rho)$ is a valid coherence measure under strictly incoherent operations and genuinely incoherent operations, it is not a valid coherence measure under incoherent operations since it does not satisfy condition $(C3)$ under incoherent operations. We now demonstrate this point by giving a counter example, which is found by trial and error.

The state in the counter example is
\begin{align}
\rho=a
\left(
    \begin{array}{ccccc}
        \frac{97}{7191} & \frac{166}{8961} & \frac{141}{3829} & \frac{210}{3449} & \frac{208}{4411}\\
        \frac{166}{8961} & \frac{311}{3585} & \frac{106}{815} & \frac{321}{8594} & \frac{158}{2091}\\
        \frac{141}{3829} & \frac{106}{815} & \frac{847}{2561} & \frac{224}{1347} & \frac{461}{2327}\\
        \frac{210}{3449} & \frac{321}{8594} & \frac{224}{1347} & \frac{333}{964} & \frac{401}{1658}\\
        \frac{208}{4411} & \frac{158}{2091} & \frac{461}{2327} & \frac{401}{1658} & \frac{561}{2509}\\
    \end{array}
\right)
\notag
\end{align}
with $a=\frac{31496~17124~91780}{31496~18253~15173}$, and the incoherent operation is
\begin{align}
\Lambda(\rho)=K_1\rho K_1^\dag+K_2\rho K_2^\dag
\end{align}
 with
\begin{align}
K_1=
\left(
    \begin{array}{ccccc}
        0 & 0 & 0 & 0 & \frac{1}{\sqrt{2}}\\
        \frac{3}{5} & \frac{4}{5} & 0 & 0 & 0\\
        0 & 0 & \frac{1}{\sqrt{2}} & 0 & 0\\
        0 & 0 & 0 & \frac{1}{\sqrt{2}} & 0\\
        0 & 0 & 0 & 0 & 0\\
    \end{array}
\right)
\nonumber
,~~K_2=
\left(
    \begin{array}{ccccc}
        0 & 0 & 0 & 0 & \frac{1}{\sqrt{2}}\\
        \frac{4}{5} & -\frac{3}{5} & 0 & 0 & 0\\
        0 & 0 & \frac{1}{\sqrt{2}} & 0 & 0\\
        0 & 0 & 0 & \frac{1}{\sqrt{2}} & 0\\
        0 & 0 & 0 & 0 & 0\\
    \end{array}
\right)
\notag.
\end{align}

In this case, there are
\begin{eqnarray}
\rho_1&&=\frac{K_{1}\rho K_{1}^{\dag}}{\mathrm{Tr}(K_{1}\rho K_{1}^{\dag})}\nonumber
\\&&=\frac{a}{p_1}
\left(
    \begin{array}{ccccc}
        \frac{561}{5018} & \frac{2046268\sqrt{2}}{46117005} & \frac{461}{4654} & \frac{401}{3316} & 0\\
        \frac{2046268\sqrt{2}}{46117005} & \frac{16718443723}{213900190125} & \frac{1968241\sqrt{2}}{31206350} & \frac{2460684\sqrt{2}}{74101765} & 0\\
        \frac{461}{4654} & \frac{1968241\sqrt{2}}{31206350} & \frac{847}{5122} & \frac{112}{1347} & 0\\
        \frac{401}{3316} & \frac{2460684\sqrt{2}}{74101765} & \frac{112}{1347} & \frac{333}{1928} & 0\\
        0 & 0 & 0 & 0 & 0\\
    \end{array}
\right)
\nonumber
\end{eqnarray}
and
\begin{eqnarray}
\rho_{2}&&=\frac{K_{2}\rho K_{2}^{\dag}}{\mathrm{Tr}(K_{2}\rho K_{2}^{\dag})}\nonumber\\&&=\frac{a}{p_2}
\left(
    \begin{array}{ccccc}
        \frac{561}{5018} & -\frac{58517\sqrt{2}}{15372335} & \frac{461}{4654} & \frac{401}{3316} & 0\\
        -\frac{58517\sqrt{2}}{15372335} & \frac{14168369681}{641700570375} & -\frac{378981\sqrt{2}}{15603175} & \frac{3897573\sqrt{2}}{296407060} & 0\\
       \frac{461}{4654} & -\frac{378981\sqrt{2}}{15603175} & \frac{847}{5122} & \frac{112}{1347} & 0\\
        \frac{401}{3316} & \frac{3897573\sqrt{2}}{296407060} & \frac{112}{1347} & \frac{333}{1928} & 0\\
        0 & 0 & 0 & 0 & 0\\
    \end{array}
\right),\nonumber
\end{eqnarray}
where
\begin{equation}
p_{1}=\mathrm{Tr}(K_{1}\rho K_{1}^{\dag})=a\frac{8279592399562440949}{15679843920215781000}\nonumber
\end{equation}
and
\begin{equation}
p_{2}=\mathrm{Tr}(K_{2}\rho K_{2}^{\dag})=a\frac{22200771412133764703}{47039531760647343000}.\nonumber
\end{equation}

Substituting $\rho$, $\rho_1$, and $\rho_2$ into $\tilde{C}_1(\rho)=\|\rho-\Delta\rho\|_{1}$, $\tilde{C}_1(\rho_1)=\|\rho-\Delta\rho_1\|_{1}$, and $\tilde{C}_1(\rho_2)=\|\rho-\Delta\rho_2\|_{1}$, respectively, we can calculate each of them, and finally we obtain, by numerical calculation,
\begin{eqnarray}
\sum_{n}p_{n}\tilde{C}_1(\rho_{n})-\tilde{C}_1(\rho)=0.0152,
\end{eqnarray}
 which violates condition $(C3)$. Therefore, $\tilde{C}_1(\rho)$ cannot be taken as a coherence measure under incoherent operations.

In passing, we point out that $\tilde{C}_1(\rho)$ must also violate condition $(C2)$, i.e., $(A2)$, under incoherent operations. This can be inferred from the fact that $\tilde{C}_1(\rho)$ satisfies conditions $(A1)$ and $(A3)$ (see the last subsection) but not a valid coherence measure under incoherent operations.

It is worth noting that $\tilde{C}_1(\rho)$ is the first functional that is a valid measure under strictly incoherent operations but violates monotonicity under incoherent operations, found so far. It fills up the gap mentioned in Ref. \cite{Yadin}.

\subsection{$C_1(\rho)$ is not a coherence measure under genuinely incoherent operations}

It has been shown that $C_1(\rho)=\min_{\sigma\in\mathcal{I}}||\rho-\sigma||_1$ is not a valid coherence measure under both incoherent operations and strictly incoherent operations \cite{Yu}. We here show that $C_1(\rho)$ is not a valid coherence measure under genuinely incoherent operations too.
To this end, we give a counter example.

Let us consider the state
\begin{align}
\rho=
\left(
    \begin{array}{ccccc}
        \frac{1}{4} & \frac{1}{4} & 0 & 0 & 0\\
        \frac{1}{4} & \frac{1}{4} & 0 & 0 & 0\\
        0 & 0 & \frac{1}{6} & \frac{1}{6} & \frac{1}{6}\\
        0 & 0 & \frac{1}{6} & \frac{1}{6} & \frac{1}{6}\\
        0 & 0 & \frac{1}{6} & \frac{1}{6} & \frac{1}{6}\\
    \end{array}
\right)
\nonumber
\end{align}
and  the operation $\Lambda(\cdot)=K_1(\cdot)K_1^\dag+K_2(\cdot)K_2^\dag$ with
\begin{align}
K_1=
\left(
    \begin{array}{ccccc}
        1 & 0 & 0 & 0 & 0\\
        0 & 1 & 0 & 0 & 0\\
        0 & 0 & 0 & 0 & 0\\
        0 & 0 & 0 & 0 & 0\\
        0 & 0 & 0 & 0 & 0\\
    \end{array}
\right)
~~\text{and}~~~~
K_2=
\left(
    \begin{array}{ccccc}
        0 & 0 & 0 & 0 & 0\\
        0 & 0 & 0 & 0 & 0\\
        0 & 0 & 1 & 0 & 0\\
        0 & 0 & 0 & 1 & 0\\
        0 & 0 & 0 & 0 & 1\\
    \end{array}
\right)\nonumber.
\end{align}
It is easy to see that $\Lambda(\cdot)$ is a genuinely incoherent operation.

In this case, there are
\begin{align}\label{Tong6}
\rho_{1}=\frac{K_{1}\rho K_{1}^{\dag}}{\mathrm{Tr}(K_{1}\rho K_{1}^{\dag})}=\frac12
\left(
    \begin{array}{ccccc}
        1 &1 & 0 & 0 & 0\\
        1 &1 & 0 & 0 & 0\\
        0 & 0 & 0 & 0 & 0\\
        0 & 0 & 0 & 0 & 0\\
        0 & 0 & 0 & 0 & 0\\
    \end{array}
\right)
\end{align}
and
\begin{align}\label{Tong7}
\rho_{2}=\frac{K_{2}\rho K_{2}^{\dag}}{\mathrm{Tr}(K_{2}\rho K_{2}^{\dag})}=\frac13
\left(
    \begin{array}{ccccc}
        0 & 0 & 0 & 0 & 0\\
        0 & 0 & 0 & 0 & 0\\
        0 & 0 & 1& 1&1\\
        0 & 0 & 1 & 1 & 1\\
        0 & 0 & 1& 1 & 1\\
    \end{array}
\right)
\nonumber\\
\end{align}
with $p_{1}=\mathrm{Tr}(K_{1}\rho K_{1}^{\dag})=\frac12$ and $p_{2}=\mathrm{Tr}(K_{2}\rho K_{2}^{\dag})=\frac12$.

By the definition of $C_1(\rho)$, we have
\begin{eqnarray}\label{Tong9}
C_1(\rho)=\min_{\sigma\in\mathcal{I}}||\rho-\sigma||_1\leq||\rho-\sigma_0||_1=1,
\end{eqnarray}
where $\sigma_0=\mathrm{diag}(\frac12,\frac12,0,0,0)$.

To calculate $C_1(\rho_1)=\min_{\sigma\in\mathcal{I}}||\rho_1-\sigma||_1$,  we explicitly write $\sigma$ as
\begin{align}\label{TongA}
\sigma=\left(
    \begin{array}{cc}
        \sigma_{00} & 0\\
       0 & \sigma_{11}\\
    \end{array}
\right)\bigoplus
\left(
    \begin{array}{ccc}
        \sigma_{22} & 0 & 0\\
         0 & \sigma_{33} & 0\\
         0 & 0 & \sigma_{44}\\
    \end{array}
\right)
\end{align}
with its diagonal elements being $\sigma_{ii}\geq 0$ and satisfying $\sum_{i=0}^4\sigma_{ii}=1$,  and rewrite $\rho_1$ as
\begin{eqnarray}\label{TongA}
\rho_1=\frac{1}{2}\left(
    \begin{array}{cc}
        1 & 1\\
      1 &1\\
    \end{array}
\right)\bigoplus
\left(
    \begin{array}{ccc}
        0 & 0 & 0\\
         0 & 0 & 0\\
         0 & 0 & 0\\
    \end{array}
\right).
\end{eqnarray}
By using the fact that $||M_1\bigoplus M_2||_1=||M_1||_1+||M_2||_1$,  we then have
\begin{eqnarray}\label{TongB}
||\rho_1-\sigma||_1&=&||\frac{1}{2}\left(
    \begin{array}{cc}
        1 & 1\\
       1 & 1\\
    \end{array}
\right)
-\left(
    \begin{array}{cc}
        \sigma_{00} & 0\\
       0 & \sigma_{11}\\
    \end{array}
\right)||_1\nonumber\\
&&+
|| \left(
    \begin{array}{ccc}
        0 & 0 & 0\\
         0 &0 & 0\\
         0 & 0 &0\\
    \end{array}
\right)
-\left(
    \begin{array}{ccc}
        \sigma_{22} & 0 & 0\\
         0 & \sigma_{33} & 0\\
         0 & 0 & \sigma_{44}\\
    \end{array}
\right)||_1\nonumber\\
&=&||\left(
    \begin{array}{cc}
        \frac{1}{2}- \sigma_{00}& \frac{1}{2}\\
       \frac{1}{2} & \frac{1}{2}- \sigma_{11}\\
    \end{array}
\right)||_1
+
|| \left(
    \begin{array}{ccc}
        \sigma_{22} & 0 & 0\\
         0 & \sigma_{33} & 0\\
         0 & 0 & \sigma_{44}\\
    \end{array}
\right)||_1\nonumber\\
&=&\sqrt{1+(\sigma_{00}-\sigma_{11})^2}+1-\sigma_{00}-\sigma_{11},
\end{eqnarray}
and we finally obtain
\begin{eqnarray}\label{TongN3}
C_1(\rho_1)&=&\min_{\sigma\in\mathcal{I}}\big[ \sqrt{1+(\sigma_{00}-\sigma_{11})^2}+(1-\sigma_{00}-\sigma_{11})\big]\nonumber\\
&=&1.
\end{eqnarray}

To calculate $C_1(\rho_2)=\min_{\sigma\in\mathcal{I}}||\rho_2-\sigma||_1$,  we rewrite $\rho_2$ as
\begin{eqnarray}\label{TongA}
\rho_2=&\left(
    \begin{array}{cc}
        0 & 0\\
      0 &0\\
    \end{array}
\right)\bigoplus
\frac{1}{3}\left(
    \begin{array}{ccc}
        1 & 1 &1\\
         1 & 1 &1\\
        1 & 1 &1\\
    \end{array}
\right).\nonumber\\
\end{eqnarray}
We then have
\begin{eqnarray}\label{TongB}
||\rho_2-\sigma||_1&=&
|| \left(
    \begin{array}{cc}
        0 & 0\\
      0 &0\\
    \end{array}
\right)
-\left(
    \begin{array}{cc}
        \sigma_{00} & 0 \\
         0 & \sigma_{11} \\
            \end{array}
\right)||_1\nonumber\\
&&+
||\frac{1}{3}\left(
    \begin{array}{ccc}
        1 & 1&1\\
       1 & 1&1\\
       1 & 1&1\\
    \end{array}
\right)
-\left(
    \begin{array}{ccc}
        \sigma_{22} & 0&0\\
       0 & \sigma_{33}&0\\
       0 & 0&\sigma_{44}\\
    \end{array}
\right)||_1.
\end{eqnarray}
To simplify the last line in Eq. (\ref{TongB}), we let $U_k=\sum_{i=0}^2\ket{i+k}\bra{i}$ with $(i+k)$ being mod(3) and $k=0,1,2$, and use the relation $||UMU^\dag||_1=||M||_1$ for any unitary operator $U$. We have
\begin{eqnarray}\label{TongN2}
&&||\frac{1}{3}\left(
    \begin{array}{ccc}
        1 & 1&1\\
       1 & 1&1\\
       1 & 1&1\\
    \end{array}
\right)
-\left(
    \begin{array}{ccc}
        \sigma_{22} & 0&0\\
       0 & \sigma_{33}&0\\
       0 & 0&\sigma_{44}\\
    \end{array}
\right)||_1\nonumber\\
&=&\frac{1}{3}\sum_{k=0}^2||U_k\left[\frac{1}{3}\left(
    \begin{array}{ccc}
        1 & 1&1\\
       1 & 1&1\\
       1 & 1&1\\
    \end{array}
\right)
-\left(
    \begin{array}{ccc}
        \sigma_{22} & 0&0\\
       0 & \sigma_{33}&0\\
       0 & 0&\sigma_{44}\\
    \end{array}
\right)\right]U_k^\dag||_1\nonumber\\
&\geq&
\frac{1}{3}||\sum_{k=0}^2 U_k\left[\frac{1}{3}\left(
    \begin{array}{ccc}
        1 & 1&1\\
       1 & 1&1\\
       1 & 1&1\\
    \end{array}
\right)
-\left(
    \begin{array}{ccc}
        \sigma_{22} & 0&0\\
       0 & \sigma_{33}&0\\
       0 & 0&\sigma_{44}\\
    \end{array}
\right)\right]U_k^\dag||_1\nonumber\\
&=&||\frac{1}{3}\left(
    \begin{array}{ccc}
        1 & 1&1\\
       1 & 1&1\\
       1 & 1&1\\
    \end{array}
\right)
-\frac{\sigma_{22}+\sigma_{33}+\sigma_{44}}{3}I||_1\nonumber\\
&=&1+\frac{\sigma_{22}+\sigma_{33}+\sigma_{44}}{3}.
\end{eqnarray}
From Eqs. (\ref{TongB}) and  (\ref{TongN2}), we obtain
\begin{eqnarray}
||\rho_2-\sigma||_1\geq\frac{4}{3}+\frac{2(\sigma_{00}+\sigma_{11})}{3}.
\end{eqnarray}
We then have
\begin{eqnarray}\label{TongN4}
C_1(\rho_2)=\min_{\sigma\in\mathcal{I}}||\rho_2-\sigma||_1 \geq\frac{4}{3}.
\end{eqnarray}
On the other hand,
\begin{eqnarray}\label{TongN5}
C_1(\rho_2)=\min_{\sigma\in\mathcal{I}}||\rho_2-\sigma||_1 \leq ||\rho_2-\Delta\rho_2||_1 =\frac{4}{3}.
\end{eqnarray}
From Eqs. (\ref{TongN4}) and  (\ref{TongN5}), we finally obtain
\begin{eqnarray}\label{TongN6}
C_1(\rho_2)=\frac{4}{3}.
\end{eqnarray}

By using the results in Eqs. (\ref{Tong9}), (\ref{TongN3}), and (\ref{TongN6}) and noting that $p_1=p_2=\frac{1}{2}$, we immediately obtain
\begin{align}
\sum_{n}p_{n}C_1(\rho_{n})-C_1(\rho)&\geq\frac12+\frac23-1=\frac16>0,
\end{align}
which violates condition $(C3)$.

Therefore, $C_1(\rho)$ cannot be taken as a coherence measure under genuinely incoherent operations.

 \begin{table*}\label{table2}
\caption{$\tilde{C}_1(\rho)$ is a valid coherence measure under strictly incoherent operations and genuinely incoherent operations, while all others, including $\tilde{C}_1(\rho)$ under incoherent operations, $\tilde{C}_{p>1}(\rho)$ under any of the three sets of free operations, and $C_p(\rho)$ under any of the three sets of free operations, cannot be taken as a coherence measure. This fills up the gaps in Table \ref{table1}.}
\begin{tabular}{cccc}
    \hline
    \hline
             &  Incoherent operations    &  Strictly incoherent operations  & Genuinely incoherent operations  \\
    \hline
    $C_1(\rho)$ & Not a coherence measure& Not a coherence measure &  Not a coherence measure \\
    $\tilde{C}_1(\rho)$ &  Not a coherence measure &  A coherence measure&  A coherence measure\\
    $C_{p>1}(\rho)$ &  Not a coherence measure& Not a coherence measure  &  Not a coherence measure \\
    $\tilde{C}_{p>1}(\rho)$ &  Not a coherence measure &  Not a coherence measure&  Not a coherence measure  \\
    \hline
    \hline
    \end{tabular}
\end{table*}

\subsection{ Neither $C_{p>1}(\rho)$ nor $\tilde{C}_{p>1}(\rho)$  is a coherence measure under genuinely incoherent operations}

Finally, we  show that neither $\tilde{C}_{p>1}(\rho)$ nor $C_{p>1}(\rho)$   fulfills condition (C3) even under genuinely incoherent operations.

Let us consider the state
\begin{align}
\rho=
\left(
    \begin{array}{cccc}
        \frac{1}{4} & 0 & \frac{1}{8} & 0\\
        0 & \frac{1}{4} & 0 & \frac{1}{8}\\
        \frac{1}{8} & 0 & \frac{1}{4} & 0\\
        0 & \frac{1}{8} & 0 & \frac{1}{4}\\
    \end{array}
\right)  \label{example state 1}
\end{align}
and the genuinely incoherent operation
\begin{equation}
\Lambda(\cdot)=K_1(\cdot)K_1^\dag+K_2(\cdot)K_2^\dag+K_3(\cdot)K_3^\dag+K_4(\cdot)K_4^\dag
\end{equation}
 with
\begin{align}
K_{1}=
\left(
    \begin{array}{cccc}
        \frac{1}{\sqrt{2}} & 0 & 0 & 0\\
        0 & 0 & 0 & 0\\
        0 & 0 & \frac{1}{\sqrt{2}} & 0\\
        0 & 0 & 0 & 0\\
    \end{array}
\right)
,~~~~
K_{2}=
\left(
    \begin{array}{cccc}
        \frac{1}{\sqrt{2}} & 0 & 0 & 0\\
        0 & 0 & 0 & 0\\
        0 & 0 & \frac{1}{\sqrt{2}} & 0\\
        0 & 0 & 0 & 0\\
    \end{array}
\right),
\nonumber\\
K_{3}=
\left(
    \begin{array}{cccc}
        0 & 0 & 0 & 0\\
        0 & \frac{1}{\sqrt{2}} & 0 & 0\\
        0 & 0 & 0 & 0\\
        0 & 0 & 0 & \frac{1}{\sqrt{2}}\\
    \end{array}
\right)
,~~~~
K_{4}=
\left(
    \begin{array}{cccc}
        0 & 0 & 0 & 0\\
        0 & \frac{1}{\sqrt{2}} & 0 & 0\\
        0 & 0 & 0 & 0\\
        0 & 0 & 0 & \frac{1}{\sqrt{2}}\\
    \end{array}
\right). \label{example operation 1}
\end{align}

In this case, there are
\begin{align}
\rho_{1}=\frac{K_1\rho K_1^{\dag}}{\Tr(K_1\rho K_1^{\dag})}=
\left(
    \begin{array}{cccc}
        \frac{1}{2} & 0 & \frac{1}{4} & 0\\
        0 & 0 & 0 & 0\\
        \frac{1}{4} & 0 & \frac{1}{2} & 0\\
        0 & 0 & 0 & 0\\
    \end{array}
\right),
\nonumber\\
\rho_{2}=\frac{K_2\rho K_2^{\dag}}{\Tr(K_2\rho K_2^{\dag})}=
\left(
    \begin{array}{cccc}
        \frac{1}{2} & 0 & \frac{1}{4} & 0\\
        0 & 0 & 0 & 0\\
        \frac{1}{4} & 0 & \frac{1}{2} & 0\\
        0 & 0 & 0 & 0\\
    \end{array}
\right),
\nonumber\\
\rho_{3}=\frac{K_3\rho K_3^{\dag}}{\Tr(K_3\rho K_3^{\dag})}=
\left(
    \begin{array}{cccc}
        0 & 0 & 0 & 0\\
        0 & \frac{1}{2} & 0 & \frac{1}{4}\\
        0 & 0 & 0 & 0\\
        0 & \frac{1}{4} & 0 & \frac{1}{2}\\
    \end{array}
\right),
\nonumber\\
\rho_{4}=\frac{K_4\rho K_4^{\dag}}{\Tr(K_4\rho K_4^{\dag})}=
\left(
    \begin{array}{cccc}
        0 & 0 & 0 & 0\\
        0 & \frac{1}{2} & 0 & \frac{1}{4}\\
        0 & 0 & 0 & 0\\
        0 & \frac{1}{4} & 0 & \frac{1}{2}\\
    \end{array}
\right),
\nonumber
\end{align}
and
\begin{align}
p_{1}=\mathrm{Tr}(K_{1}\rho K_{1}^{\dag})=\frac14,
~~
p_{2}=\mathrm{Tr}(K_{2}\rho K_{2}^{\dag})=\frac14,
\nonumber\\
p_{3}=\mathrm{Tr}(K_{3}\rho K_{3}^{\dag})=\frac14,
~~
p_{4}=\mathrm{Tr}(K_{4}\rho K_{4}^{\dag})=\frac14.\nonumber
\end{align}

Substituting $\rho$ and $\rho_n$ into  $\tilde{C}_{p}(\rho)$ and  $\tilde{C}_{p}(\rho_n)$, respectively,  we can obtain
\begin{eqnarray}
\tilde{C}_p(\rho)=2^{\frac{2}{p}-3} \nonumber\\
\end{eqnarray}
and
\begin{eqnarray}
\tilde{C}_p(\rho_{1})=\tilde{C}_p(\rho_{2})=\tilde{C}_p(\rho_{3})=\tilde{C}_p(\rho_{4})=2^{\frac{1}{p}-2},\nonumber\\
\end{eqnarray}
hence
\begin{eqnarray}
\sum_{n}p_{n}\tilde{C}_p(\rho_{n})-\tilde{C}_p(\rho)=2^{\frac{1}{p}-2}(1-2^{\frac{1}{p}-1})>0 \label{result1}
\end{eqnarray}
for $p>1$.
Therefore, $\tilde{C}_{p>1}(\rho)$ violates condition $(C3)$ under genuinely incoherent operations.

By using the same counter example defined by Eqs. (\ref{example state 1})-(\ref{example operation 1}), with the help of the result for $\tilde{C}_{p>1}(\rho)$, we can demonstrate  that $C_{p>1}(\rho)$ also violates condition $(C3)$ under genuinely incoherent operations.

To this end, we need to calculate $C_p(\rho_n)=\min_{\sigma\in\mathcal{I}}||\rho_n-\sigma||_p$, ~$n=1,2,3,4$. By letting $U_1=\ket{0}\bra{0}+\ket{1}\bra{2}+\ket{2}\bra{1}+\ket{3}\bra{3}$ and $U_2=\ket{0}\bra{1}+\ket{1}\bra{3}+\ket{2}\bra{2}+\ket{3}\bra{0}$, we can see that $U_1\rho_1U_1^\dag=U_1\rho_2U_1^\dag=U_2\rho_3U_2^\dag=U_2\rho_4U_2^\dag=\rho_0$,
where $\rho_0=\left(
    \begin{array}{cc}
        \frac{1}{2} & \frac{1}{4}\\
      \frac{1}{4} & \frac{1}{2}\\
    \end{array}
\right)\bigoplus
\left(
    \begin{array}{cc}
        0 &0\\
         0 &0\\
        \end{array}
\right)$.
Since $\min_{\sigma\in\mathcal{I}}||\rho_0-U_k\sigma U_k^\dag||_p=\min_{\sigma\in\mathcal{I}}||\rho_0-\sigma||_p$ for $k=1,2$, there is $C_p(\rho_n)=C_p(\rho_0)=\min_{\sigma\in\mathcal{I}}||\rho_0-\sigma||_p$, $n=1,2,3,4$.
By directly calculating the eigenvalues of $(\rho_0-\sigma)$ with $\sigma=\text{diag}\{\sigma_{00},\sigma_{11},\sigma_{22},\sigma_{33}\}$ , we have
\begin{eqnarray}\label{min}
C_p(\rho_n)=C_p(\rho_0)=\min_{\sigma\in\mathcal{I}}(|\uplambda_1|^p+|\uplambda_2|^p
+|\uplambda_3|^p+|\uplambda_4|^p)^\frac{1}{p},
\end{eqnarray}
where $\uplambda_1=\frac{1}{2}[1-\sigma_{00}-\sigma_{11}+\sqrt{\frac{1}{4}+(\sigma_{00}-\sigma_{11})^2}]$, ~ $\uplambda_2=\frac{1}{2}[1-\sigma_{00}-\sigma_{11}-\sqrt{\frac{1}{4}+(\sigma_{00}-\sigma_{11})^2}]$, ~$\uplambda_3=-\sigma_{22}$, and $\uplambda_4=-\sigma_{33}$.
Since $|\uplambda_1|^p+|\uplambda_2|^p\geq 2 \left(\frac{1}{2}\sqrt{\frac{1}{4}+(\sigma_{00}-\sigma_{11})^2}\right)^p \geq 2^{1-2p}$  \cite{TongA}, there is
\begin{eqnarray}\label{Tong12a}
C_p(\rho_n)=\min_{\sigma\in\mathcal{I}}(|\uplambda_1|^p+|\uplambda_2|^p
+|\uplambda_3|^p+|\uplambda_4|^p)^\frac{1}{p}\geq 2^{\frac{1}{p}-2}.
\end{eqnarray}
On the other hand,
\begin{eqnarray}\label{Tong12b}
C_p(\rho_n)&=&\min_{\sigma\in\mathcal{I}}(|\uplambda_1|^p+|\uplambda_2|^p
+|\uplambda_3|^p+|\uplambda_4|^p)^\frac{1}{p}\nonumber\\
&\leq& (|\uplambda_1|^p+|\uplambda_2|^p
+|\uplambda_3|^p+|\uplambda_4|^p)^\frac{1}{p}\big|_{\sigma_{00}=\sigma_{11}=\frac{1}{2},\sigma_{22}=\sigma_{33}=0}\nonumber\\
&=& 2^{\frac{1}{p}-2}.
\end{eqnarray}
From Eqs. (\ref{Tong12a}) and (\ref{Tong12b}), we immediately obtain
\begin{eqnarray}
C_p(\rho_n)= 2^{\frac1p-2}=\tilde{C}_p(\rho_n).
\end{eqnarray}

Also, by the definitions of the two Schatten-$p$-norm-based functionals, there is $C_p(\rho)\leq\tilde{C}_p(\rho)$. We then obtain
\begin{align}
\sum_{n}p_{n}C_p(\rho_{n})-C_p(\rho)\geq\sum_{n}p_{n}\tilde{C}_p(\rho_{n})-\tilde{C}_p(\rho)>0,
\end{align}
which implies that ${C}_{p>1}(\rho)$ violates condition $(C3)$ since $\tilde{C}_{p>1}(\rho)$ does. Hence, neither $C_{p>1}(\rho)$ nor $\tilde{C}_{p>1}(\rho)$ fulfills the monotonicity condition under genuinely incoherent operations, which resolves the open question raised in Ref. \cite{Vicente}.

Since there is $\mathcal{S}_{GIO}\subset \mathcal{S}_{SIO} \subset \mathcal{S}_{IO}$, we then immediately obtain that neither $C_{p>1}(\rho)$ nor $\tilde{C}_{p>1}(\rho)$ is a valid coherence measure under incoherent operations, strictly incoherent operations, and genuinely incoherent operations. Obviously, all the Schatten-$p$-norm-based functionals discussed in the paper are not valid coherence measures under maximally incoherent operations since the set of maximally incoherent operations contains incoherent operations as a subset.

\section{Conclusions}

So far, we have resolved all the open questions on whether the two classes of Schatten-$p$-norm-based functionals $C_p(\rho)=\min_{\sigma\in\mathcal{I}}||\rho-\sigma||_p$ and $ \tilde{C}_p(\rho)= \|\rho-\Delta\rho\|_{p}$ with $p\geq 1$ are valid coherence measures under incoherent operations, strictly incoherent operations, and genuinely incoherent operations, filling up the gaps in Table \ref{table1}. Our results show that only $\tilde{C}_1(\rho)$ is a valid coherence measure under strictly incoherent operations and genuinely incoherent operations, while all others, including $\tilde{C}_1(\rho)$ under incoherent operations, $\tilde{C}_{p>1}(\rho)$ under any of the three sets of free operations and $C_p(\rho)$ under any of the three sets of free operations, cannot be taken as a coherence measure, as listed in Table II.

\begin{acknowledgments}

We acknowledges support from the National Natural Science Foundation of China through Grant No. 11775129 and the China
Postdoctoral Science Foundation Grant No. 2019M660841.

\end{acknowledgments}


\begin{thebibliography}{25}
\bibitem{TongA}  For $p>1$, there is always $|a+b|^p+|a-b|^p\geq 2 a^p\geq 2 b^p$  for any $a\geq b \geq 0$, since both $f(x)=x^p$ and $f^\prime(x)=px^{p-1}$ are increasing functions with respect to $x\geq0$. Now, we let  $a=\frac{1}{2}(1-\sigma_{00}-\sigma_{11})$ and $b=\frac{1}{2}\sqrt{\frac{1}{4}+(\sigma_{00}-\sigma_{11})^2}$ if  $(1-\sigma_{00}-\sigma_{11})\geq \sqrt{\frac{1}{4}+(\sigma_{00}-\sigma_{11})^2}$, and let  $b=\frac{1}{2}(1-\sigma_{00}-\sigma_{11})$ and $a=\frac{1}{2}\sqrt{\frac{1}{4}+(\sigma_{00}-\sigma_{11})^2}$ if  $(1-\sigma_{00}-\sigma_{11})\leq \sqrt{\frac{1}{4}+(\sigma_{00}-\sigma_{11})^2}$. In any case, we can immediately obtain $|\uplambda_1|^p+|\uplambda_2|^p\geq 2 \left(\frac{1}{2}\sqrt{\frac{1}{4}+(\sigma_{00}-\sigma_{11})^2}\right)^p \geq 2^{1-2p}$.
\end{thebibliography}
\end{document}